\documentclass[conference]{IEEEtran}
\IEEEoverridecommandlockouts
\usepackage{subfigure}
\usepackage{algorithm}
\usepackage{algpseudocode}
\usepackage{cite}
\usepackage{amsmath,amssymb,amsfonts}
\usepackage{graphicx}
\usepackage{textcomp}
\usepackage{xcolor}
\usepackage{balance}
\def\BibTeX{{\rm B\kern-.05em{\sc i\kern-.025em b}\kern-.08em
    T\kern-.1667em\lower.7ex\hbox{E}\kern-.125emX}}
\begin{document}

\title{PRISM: Processing-In-Memory Sparse MTTKRP for Tensor Decomposition Acceleration}

\author{\IEEEauthorblockN{Daniel Pacheco}
\IEEEauthorblockA{\textit{INESC-ID} \\
\textit{Instituto Superior Técnico}\\
\textit{Universidade de Lisboa}\\
Lisboa, Portugal \\
daniel.pacheco@inesc-id.pt}
\and
\IEEEauthorblockN{Leonel Sousa}
\IEEEauthorblockA{\textit{INESC-ID} \\
\textit{Instituto Superior Técnico}\\
\textit{Universidade de Lisboa}\\
Lisboa, Portugal \\
las@inesc-id.pt}
\and
\IEEEauthorblockN{Aleksandar Ilic}
\IEEEauthorblockA{\textit{INESC-ID} \\
\textit{Instituto Superior Técnico}\\
\textit{Universidade de Lisboa}\\
Lisboa, Portugal \\
aleksandar.ilic@inesc-id.pt}
}

\maketitle

\begin{abstract}
Sparse tensors are the most used representation of sparse multidimensional data. Operations that decompose them, selecting their most important features while reducing their dimension, have become prevalent procedures in machine learning. One of the most used tensor decomposition algorithms is the Alternating Least Squares Canonical Polyadic Decomposition (CP-ALS), where the most time-consuming operation is the Sparse Matricized Tensor Times Khatri-Rao Product (spMTTKRP). This operation is strongly memory-bound, making it hard to implement efficiently on general-purpose processors. This work proposes PRISM, the first approach to tackle this operation using Processing-In-Memory (PIM) technology. We extensively characterize different partitioning strategies, number formats, and kernel optimizations that efficiently adapt this operation to UPMEM PIM, which is further boosted by heterogeneous collaboration with the CPU. The experimental results show that the proposed PIM-based and heterogeneous approaches achieve up to 2.37 and 2.64 times speedup compared to state-of-the-art CPU implementations, respectively. However, the UPMEM distributed memory system can significantly hinder performance on certain workloads. Nonetheless, the efficiency of resource consumption for this approach, measured by peak performance fraction usage, is significantly higher than for both CPU and GPU.\looseness=-1
\end{abstract}


\section{Introduction}

Machine learning methods often require high amounts of multi-dimensional data that, when used directly, can lead to low performance, both in execution time and accuracy. This data, often represented as tensors, can be decomposed by selecting its most important features, creating a much smaller data structure that retains the most relevant information. By removing redundant elements, this technique can be used to increase the accuracy of machine learning methods \cite{nlp, sp}. Moreover, it also provides data compression by reducing the size of the tensor while maintaining its information.
One of the main tensor decomposition algorithms is the Canonical Polyadic Decomposition (CPD) \cite{CP} via alternating least squares (CP-ALS), for which the most time-consuming operation is the Matricized Tensor Times Khatri-Rao Product (MTTKRP) \cite{HiCOO, ALTO, FLYCOO_FPGA, Dynasor, cp1, cp2, GPU_spMTTKRP, BLCO}.\looseness=-1

When applied to a sparse tensor, the MTTKRP becomes a Sparse Matricized Tensor Times Khatri-Rao Product (spMTTKRP). The spMTTKRP is a heavily memory-bound operation, as it only uses fundamental arithmetic operations, namely multiplications and sums, but performs multiple non-contiguous memory accesses. This reduces the effectiveness of caches, making this operation hard to implement efficiently on general-purpose processors. Several publications investigate algorithms and implementations on CPU \cite{HiCOO, ALTO, Dynasor, cp1} and GPU \cite{cp2, GPU_spMTTKRP, BLCO}, mitigating this issue by either designing new tensor formats that reduce the number of memory transfers performed or using mapping strategies that maximize data reusability. However, one alternative that has not yet been explored in the literature is using hardware that is better suited for memory-bound applications, particularly Processing In Memory (PIM). This technology moves computation closer to data by equipping memory chips with processing capabilities.\looseness=-1

One of the most used commercially available systems that implements this technology is UPMEM PIM, which integrates DRAM Processing Units (DPUs) into the DRAM banks. These units can perform simple arithmetic operations while benefiting from memory accesses that provide both low latency and high bandwidth. However, exploring the PIM capabilities of this device comes with its own set of challenges, as it is composed of a distributed memory system and does not natively support floating-point arithmetic.\looseness=-1

This work proposes PRISM, the first method for performing spMTTKRP using UPMEM PIM. To tackle the challenges mentioned above, PRISM creates a new spMTTKRP design that adopts a new tensor format to operate in this memory system, and also provides a technique for substituting the usual floating-point representation with fixed-point arithmetic. This work also explores a heterogeneous approach that utilizes both CPU and UPMEM PIM, leveraging otherwise idle resources. The full CP-ALS algorithm is also developed, leaving the other operations performed in CP-ALS to the CPU. This allows analysis of how the strategies used impact the convergence of decomposition and the trade-off between performance and accuracy.\looseness=-1



The key contributions of this work are the following:
\begin{itemize}
\item \textbf{PIM-oriented mapping:} This work investigates a partitioning system that allows any tensor and factor matrices to be mapped onto the distributed memory system of UPMEM PIM, supported by the development of a novel sparse tensor representation. It also creates a partitioning algorithm that aims at fully exploiting the capabilities of this hardware, while optimizing its performance for larger workloads.\looseness=-1

\item \textbf{Number formats:} To allow efficient usage of UPMEM, which does not natively support floating-point arithmetic, this work presents an spMTTKRP approach using multiple fixed-point formats. We also conduct an extensive characterization of the performance-accuracy tradeoffs across a range of different data formats and tensor characteristics.\looseness=-1

\item \textbf{Optimized PIM-based kernels:} Several kernel optimization strategies are explored in this paper, such as tasklets, sequential readers, and lock removal, taking advantage of the inherent characteristics of spMTTKRP and CP-ALS to maximize throughput.\looseness=-1

\item \textbf{Heterogeneous UPMEM+CPU processing:} To use all the available processing power and further improve the performance, we split the spMTTKRP operation between UPMEM PIM and CPU. A well-suited load-balancing algorithm uses the partitioning system developed to distribute the work between both devices. This distribution is done according to the characteristics of each partition, where partitions sent to PIM are the ones that best fit its features.\looseness=-1

\end{itemize}
PRISM is thoroughly evaluated and analyzed on multiple workloads to characterize the benefits of this approach. For optimal cases, the PIM-only and heterogeneous approaches can reach speedups of 2.37 times and 2.64 times, respectively, over the state-of-the-art CPU implementations. Less performant workloads are also investigated to determine the limitations of the architecture. However, in all workloads, the fraction between the obtained and the maximum performance in PIM is much higher than in the CPU or GPU, indicating a much more efficient usage of the PIM device.\looseness=-1


\section{Background}

This section summarizes the concepts required to understand the remainder of this work, namely, Canonical Polyadic Decomposition (CPD), UPMEM PIM architecture, and element-wise spMTTKRP.\looseness=-1

\subsection{Canonical Polyadic Decomposition} \label{cpd_explanation}

A tensor is defined as a multi-dimensional data structure with the number of dimensions not bounded. Each dimension of a tensor is called a mode. A mode-N tensor is commonly represented by $\mathcal{X} \in \mathrm{R}^{I_0 \times ... \times I_{N-1}}$. When a mode-N tensor can be represented as an outer product of N vectors, it is called a rank-one tensor.\looseness=-1

The CPD aims to approximate a tensor to a sum of R rank-one tensors, where R is the decomposition rank. 
It can be defined as
\begin{equation}
    \mathcal{X}\approx \sum_{r=1}^R \boldsymbol{b}_r^0 \circ \boldsymbol{b}_r^1 \circ \cdots \boldsymbol{b}_r^{N-1},
\end{equation}
where $\boldsymbol{b}_r^0 \circ \boldsymbol{b}_r^1 \circ \cdots \boldsymbol{b}_r^{N-1}$ represents the vector outer product (each generating rank-one tensor), and $\boldsymbol{b}^n_r\in \mathbb{R}^{I_n}, \forall $ $ 0{\leq}n{<}N$, $\forall $ $ 1{\leq}r{\leq}R$.

To find the CPD of a tensor, the most used method is the Alternating Least Squares (CP-ALS) \cite{CP_ALS, CP_ALS2, CP_ALS3}, which is shown in Algorithm \ref{alg:cp} for a mode-3 tensor.\looseness=-1

\begin{algorithm}\small
\caption{Canonical Polyadic Decomposition}\label{alg:cp}
\includegraphics[width=0.9\linewidth]{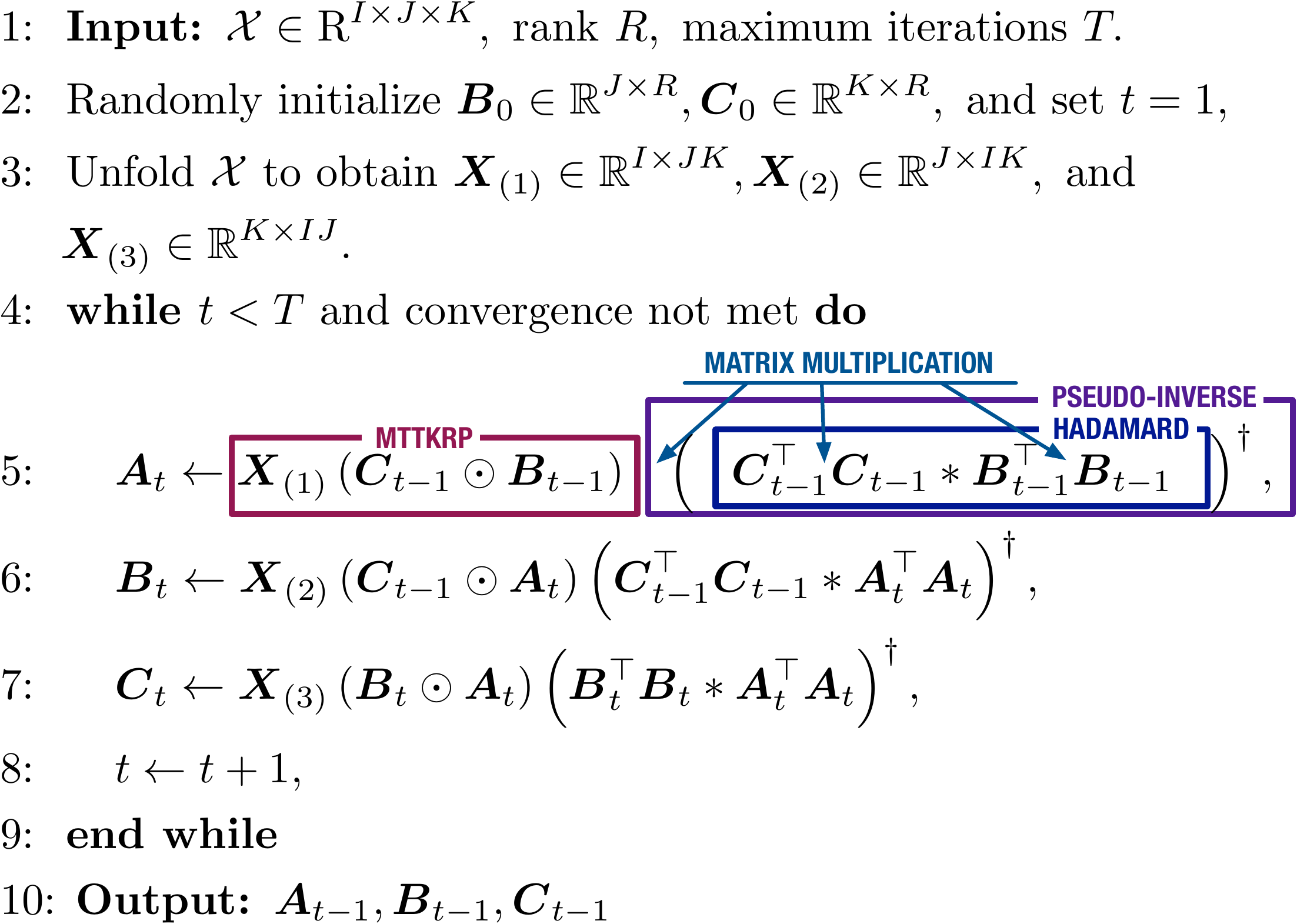}
\end{algorithm}

In the CP-ALS decomposition, each iteration performs multiple matrix multiplications, matrix transposes, Hadamard products, a matrix pseudo-inverse, and a Matricized Tensor Times Khatri-Rao Product (MTTKRP) for each tensor mode. The matrices $\boldsymbol{A}$, $\boldsymbol{B}$, and $\boldsymbol{C}$ represent the factor matrices, which contain the vectors resulting from the decomposition as their columns, and have as many rows as the decomposition rank. For example, $\boldsymbol{A}=[\boldsymbol{b}^0_1, \boldsymbol{b}^0_2, \cdots, \boldsymbol{b}^0_{R}]$. It is also common to perform a normalization step on the factor matrices at each step \cite{parti_paper}, where the normalization is performed independently for each factor matrix rank, using either L-2 or L-infinity norm. The decomposition rank is often lower than 100, while the tensor dimensions can be in the order of millions, which, coupled with the tensor having more dimensions than a matrix, makes the tensor much larger than the factor matrices and, consequently, MTTKRP the performance bottleneck of this decomposition. In this work, the focus is on performing sparse tensor decomposition, where the tensor is sparse, but the factor matrices remain dense.\looseness=-1

\subsection{Element-wise spMTTKRP} \label{elementwise_spmttkrp}

\begin{figure}[h]
\centering
\includegraphics[width=\linewidth, trim={0 0cm 0 0cm},clip]{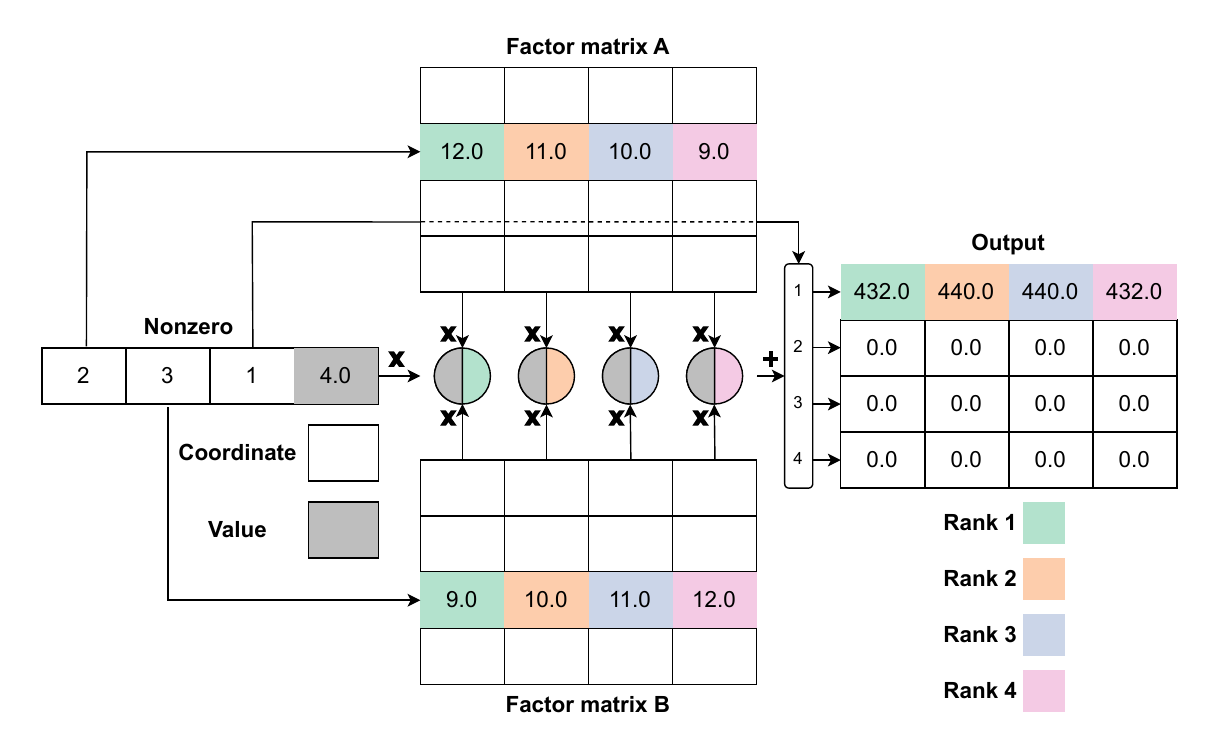}
\caption{Representation of element-wise MTTKRP}
\label{elementwise_fig}
\vspace{-10pt}
\end{figure}

The MTTKRP operation, when applied to sparse tensors, is performed element-wise, taking each nonzero and computing all its partial results, before performing the final reduction. Figure \ref{elementwise_fig} illustrates how the partial results of a nonzero are calculated for a mode-3 spMTTKRP (line 7 of Algorithm \ref{alg:cp}).\looseness=-1

On a mode-3 spMTTKRP operation, factor matrices A and B are inputs associated with the first and second tensor dimensions, respectively. This means that the first nonzero coordinate will index the row of matrix A to compute, and the second coordinate will index the row of matrix B. Then, a Hadamard product is performed between these rows, and the result is multiplied by the nonzero value, creating the nonzero partial result. Finally, this partial result is added to the output row indexed by the nonzero's third coordinate, as it is the only coordinate that did not index any factor matrix. The number of columns, both of the factor matrices and output, is the same as the decomposition rank, and each column of the result is associated with a different rank. Since there are no dependencies between columns of the partial result, each rank can be computed independently.\looseness=-1
\subsection{UPMEM PIM}

\begin{figure}[b]
\vspace{-10pt}
\centering
\includegraphics[width=0.7\linewidth]{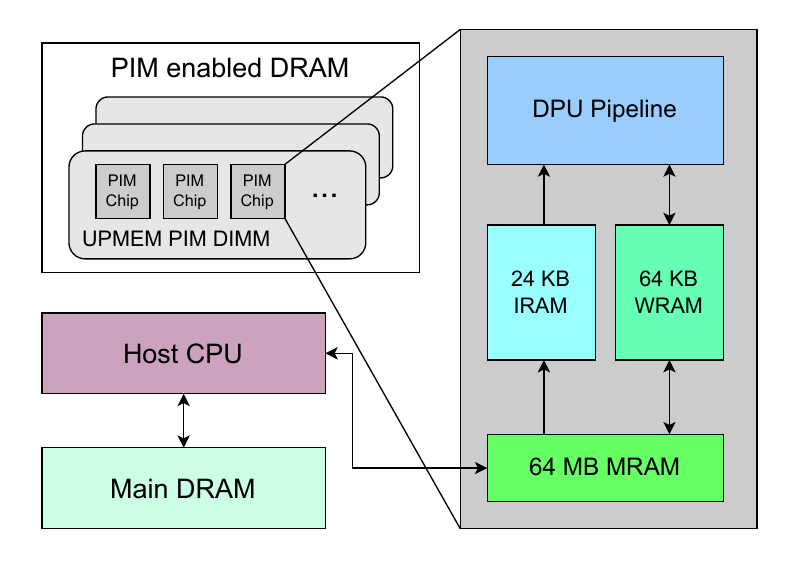}
\caption{Representation of the UPMEM PIM architecture}
\label{upmem_fig}
\end{figure}
The UPMEM PIM system, presented in Figure \ref{upmem_fig}, incorporates the host CPU with standard main memory and UPMEM PIM modules. An UPMEM PIM module includes a standard DDR4-2400 DIMM with 2 ranks. Each rank contains 64 PIM cores called DRAM Processing Units (DPUs). The current UPMEM PIM system has 20 double-rank PIM DIMMs with 2560 DPUs. Each DPU has exclusive access to a 24-KB instruction memory, called IRAM, a 64-KB scratchpad memory, called WRAM, and a 64-MB DRAM bank, called MRAM. This means there is no shared memory between DPUs or access to remote memory, so DPUs cannot communicate directly. A DPU is a multithreaded in-order 32-bit RISC core that runs at up to 500 MHz. The DPU has 24 hardware threads, each with 24 32-bit general-purpose registers. The DPU pipeline has 14 stages, and a minimum of eleven tasklets is required to fully utilize it and achieve maximum performance. The DPU natively supports single-cycle 8x8-bit multiplication and single-cycle 32-bit integer addition and subtraction. Arithmetic operations with other data types must be performed based on software, requiring more clock cycles. Floating-point arithmetic is, therefore, especially challenging for this hardware and should be avoided.\looseness=-1

\section{Related Work}

The importance of CP-ALS, coupled with the difficulty in efficiently implementing its main operation, the MTTKRP, has led multiple works to provide more efficient approaches to this operation on CPU \cite{ALTO,Dynasor,HiCOO}, GPU \cite{BLCO,GPU_spMTTKRP}, and even FPGA \cite{FLYCOO_FPGA}. These approaches use novel sparse tensor formats to increase locality and improve workload distribution with minimal synchronization costs.\looseness=-1

In \cite{ALTO}, a new sparse tensor format, Adaptive Linearized Tensor Order (ALTO), was developed to optimize spMTTKRP on multicore CPU architectures. In this format, every nonzero element is represented by its value and position, which is a linearization of its coordinates, and then ordered according to that position. This ordering improves locality for the factor matrices of spMTTKRP on all modes, increasing cache hits. Another benefit of this format is that it takes reduced memory space when compared to other formats, and only requires one memory access to get all the coordinates of a single nonzero. It can also be partitioned at any granularity among threads, improving workload balance. However, multiple bitwise operations are required to obtain the coordinates of each nonzero from its position, which can become a significant overhead, especially on devices that do not support bit manipulation operations natively. This format is adapted to GPUs in \cite{BLCO} with the Blocked Linearized Coordinate (BLCO) format, which re-arranges the bits in the linearized indexes before sending them to the GPU, allowing for a more performant extraction of the original coordinates.\looseness=-1

In \cite{HiCOO}, one of the most well-known formats for spMTTKRP was developed: HiCOO. This format groups nonzeros into blocks with a pre-defined size that must be a power of two. Then, for each block with at least one element, its coordinates are stored along with the relative coordinates of its nonzeros and their value. The blocks and
nonzeros within each block are sorted using Z-Morton order \cite{Z-Morton}. This format exploits data locality on all tensor modes, optimizing cache accesses, but its workload granularity is limited by the block size, which, for some tensors, can cause workload imbalance.\looseness=-1

In \cite{FLYCOO_FPGA}, another new format, FLYCOO, was developed for computing spMTTKRP. This format was first introduced on FPGAs, but later adapted to multicore CPU architectures in \cite{Dynasor} and GPU architectures in \cite{GPU_spMTTKRP}. The FLYCOO format assigns each nonzero to multiple tensor partitions, one for each mode, called shards. Then, the shard IDs are embedded into the elements. To form these shards, Z-Morton ordering is used to improve data locality in the elements within each shard. This format allows a finer workload granularity while maintaining data locality and completely avoiding write conflicts. However, the tensor must be reordered for each spMTTKRP mode, which can only be done efficiently by using enough memory to store two copies of the tensor.\looseness=-1

Regarding the usage of PIM and, specifically, UPMEM PIM, many works have been done for multiple applications, such as wavefront algorithm \cite{WFA}, sequence alignment \cite{sa}, join algorithm \cite{PIM_Join}, RNA sequence quantification \cite{rna_pim}, and sparse matrix-vector multiplication \cite{SparseP}. These works have shown the capabilities of this device, especially in algorithms that are heavily memory-bound. However, the research reported in this paper is a pioneer in utilizing PIM for tensor decomposition. This application, despite falling into the linear algebra category with sparse matrix-vector multiplication, tackles larger data structures, both in number and size of dimensions, proving to be a more complex challenge.\looseness=-1

\section{PRISM: Design and optimization strategies}

The main step in performing CPD is the spMTTKRP operation. Its efficient processing using UPMEM PIM brings to practice a new set of challenges, since there is no global memory, each DPU possesses a limited amount of memory, and there is no communication between DPUs.\looseness=-1

The tensor formats used in the state-of-the-art CPU and GPU approaches \cite{ALTO,BLCO,HiCOO,Dynasor,GPU_spMTTKRP} allow partitioning the nonzeros to maximize parallelism and avoid write-conflicts, but never consider partitioning the factor matrices, as this is not required on the targeted devices. These formats would, therefore, require each DPU to contain a copy of the factor matrices. This significantly limits scalability since the memory required for the factor matrices will exhaust the DPU memory for most common workloads. Therefore, we designed a new tensor format that allows not only the partitioning of the nonzeros but also the factor matrices.\looseness=-1

To enable workload distribution across decomposition ranks, tensor dimensions, and nonzeros, this work characterizes three partitioning dimensions. When coupled with the novel tensor format, they can map this operation to any number of UPMEM PIM DPUs. This partitioning strategy, although designed specifically for UPMEM PIM, can be expanded to other systems where communication between processing cores is limited.\looseness=-1

\subsection{Proposed tensor format}
\begin{figure}[t]
     \centering
     \subfigure[COO format]{\label{coo_new}
     \includegraphics[scale=0.45, trim={0 0.5cm 0 0.5cm},clip]{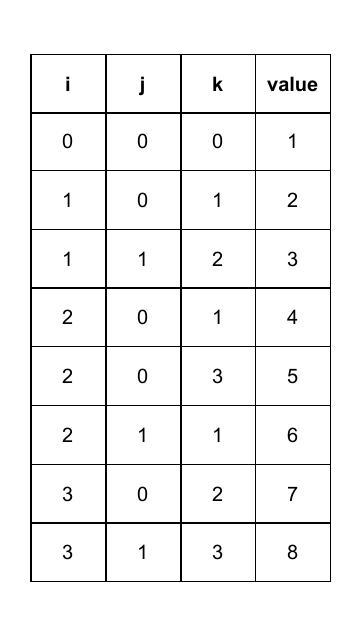}}
        \qquad
      \subfigure[Proposed format]{\label{my_tensor}
     \includegraphics[scale=0.45, trim={0 0.5cm 0 0.5cm},clip]{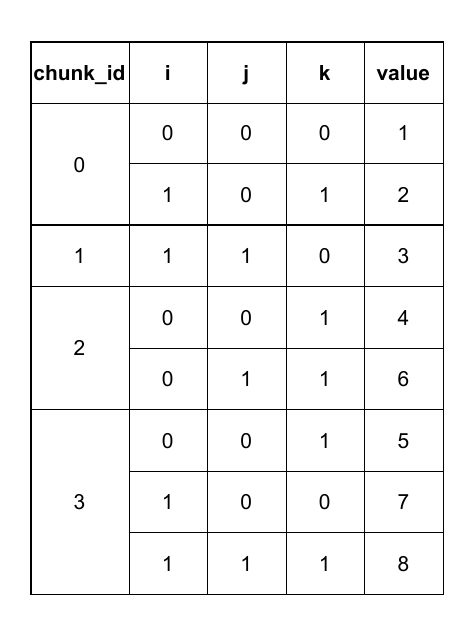}}
        \caption{COO and the proposed representation of a tensor}
        \label{fig:proposed_tensor}
        \vspace{-10pt}
\end{figure}

The proposed tensor format splits the tensor across one or multiple dimensions, creating tensor chunks with pre-defined and equal sizes. Each nonzero is inserted into the respective chunk, and its coordinates become relative to that chunk. This is represented by a two-dimensional structure where the first dimension represents the tensor chunk index and the second contains the values and relative coordinates of each nonzero. For example, for the tensor with dimensions 4x2x4  depicted in Figure \ref{fig:proposed_tensor}, when the size of each chunk is defined as 2x2x2, we create four tensor chunks to hold, respectively, elements (0:1,0:1,0:1), (0:1,0:1,2:3), (2:3,0:1,0:1), and (2:3,0:1,2:3). Then, each nonzero is inserted in the corresponding chunk, and its coordinates become relative to that chunk.\looseness=-1

This structure can be created using chunks of any size. The size of the tensor chunks in each dimension determines the required data from the corresponding factor matrix. Therefore, this format allows for the partitioning of factor matrix data, ensuring it can fit in the DPU. This format also allows values inserted into the same chunk to be transferred to the same DPU efficiently, as they will be contiguous in memory. Another benefit of this new format is that each DPU can interpret the received chunk as a complete tensor, computing each chunk independently.\looseness=-1

\subsection{Hierarchical partitioning} \label{Partitioning}





The developed approach aims to provide a flexible partitioning strategy while still allowing each processing unit to compute its own set of partial results independently. To do so, the partitioning process considers three partitioning dimensions: \textit{i)} rank partitioning, \textit{ii)} dimension size partitioning, and \textit{iii)} nonzero partitioning.\looseness=-1

\textbf{Rank partitioning} consists of sending different ranks of the factor matrices to different DPUs. This is the first level at which the partitioning is done because the computation of different ranks is completely independent, not requiring any sum reduction to be performed. 
Additionally, the only data that has to be replicated among different DPUs is the tensor data, since processing units working on different ranks will still require the same nonzeros, but different sections of the factor matrices. This is highly efficient because when performing tensor decomposition, the factor matrices change, but the tensor is always static. Therefore, the tensor data can be transferred to the processing units only once and then kept between spMTTKRP iterations. Regarding the memory usage of each DPU, since the computation of each rank only requires the corresponding rank of the factor matrices, this partitioning dimension reduces the amount of factor matrix data required per DPU.\looseness=-1

\begin{figure}[t]
\centering
\includegraphics[width=0.72\linewidth,trim={0cm 0.8cm 0 0.5cm},clip]{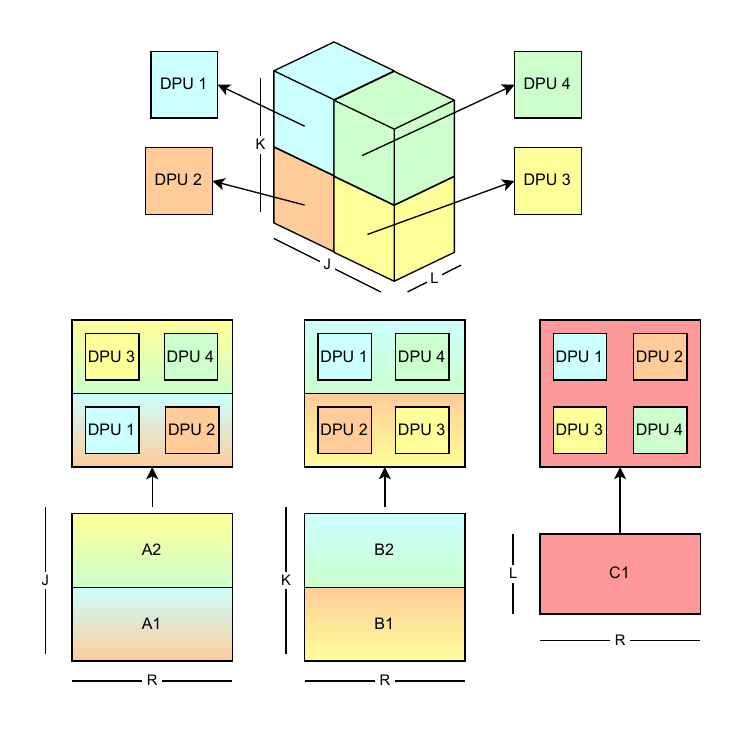}
\caption{Representation of dimension size partitioning}
\label{dim_size partition}
\vspace{-10pt}
\end{figure}

\textbf{Dimension size partitioning}, represented in Figure \ref{dim_size partition}, is tightly coupled with the tensor format used and consists of sending different tensor chunks to different DPUs. By splitting the tensor across its dimensions, this partitioning reduces the factor matrix elements required by each DPU. This happens because, as represented in Figure \ref{elementwise_fig}, the coordinates of a nonzero are used directly to index the factor matrices. Since each chunk only represents a certain set of coordinates, only the corresponding elements of the factor matrix are necessary for the DPU. Therefore, this partitioning dimension limits the factor matrix data required per DPU. However, since the number of nonzeros per chunk is not bound, it does not directly restrict the amount of tensor data necessary. The main drawback of this partitioning is that it provokes factor matrix data replication.\looseness=-1

Figure \ref{dim_size partition} exemplifies this by depicting a tensor that is split in half in both dimensions J and K, leading to an identical split of matrices A and B. Each half of these matrices is assigned to two DPUs. Matrix C, since it is not partitioned, is assigned to all DPUs. If the dimension associated with each matrix is an input dimension, this repeated assignment will lead to multiple matrix copies being sent to the DPUs. If it is on the output dimension, it will lead to multiple partial results for the same output position, requiring a sum reduction.\looseness=-1

Finally, \textbf{nonzero partitioning} consists of distributing the nonzeros of a tensor chunk among different DPUs. This is necessary when the number of nonzeros assigned to a chunk is larger than the amount a DPU can store. This partitioning dimension can, therefore, limit the amount of tensor data required by the DPUs, but has no impact on factor matrix data. When multiple DPUs are assigned different nonzeros of the same tensor chunk, both the required factor matrix data and the targeted output elements will be the same, as these are both directly related to the tensor chunk. This will require replication of factor matrix data on all input dimensions, along with a sum reduction of the partial results of the DPUs for any output dimension, generating more data replication than dimension size partitioning.\looseness=-1

Regarding the partitioning granularity, a finer granularity allows for the usage of more DPUs, which reduces the kernel time. However, performing dimension size and nonzero partitioning increases the amount of data replicated and, consequently, the number of memory transfers and sum reductions necessary. This is particularly concerning as the tensor and factor matrices grow.\looseness=-1

Given the referred properties, to optimize the design for large tensors, which are the most challenging for state-of-the-art approaches, the proposed partitioning scheme minimizes dimension size and nonzero partitioning. As analyzed before, rank partitioning does not require factor matrix data replication, so it is the partitioning technique favored in the proposed approach. Regarding the two other dimensions, dimension size partitioning is preferable to nonzero partitioning. However, completely avoiding nonzero partitioning may not be optimal. This happens because in denser areas of the tensor, the chunk size must be reduced to ensure all nonzeros fit in a DPU. However, in sparser areas of the tensor, these smaller chunks will have fewer nonzeros than the amount that can fit in a DPU. This means that the dimension size partitioning done in these areas is higher than necessary, which may outweigh the benefits of avoiding nonzero partitioning.\looseness=-1

When balancing dimension size and nonzero partitioning, it is important to understand how these partitioning dimensions affect each other. As stated before, dimension size partitioning limits the factor matrix required by each DPU, while nonzero partitioning restricts the required tensor data. Since both structures share the DPU memory, increasing the partition size in one of these dimensions, enlarging the corresponding memory usage, will require a decreased partition size in the other.\looseness=-1

\begin{figure}[t]
\centering
\includegraphics[width=0.77\linewidth,trim={0 0.5cm 0 0.5cm},clip]{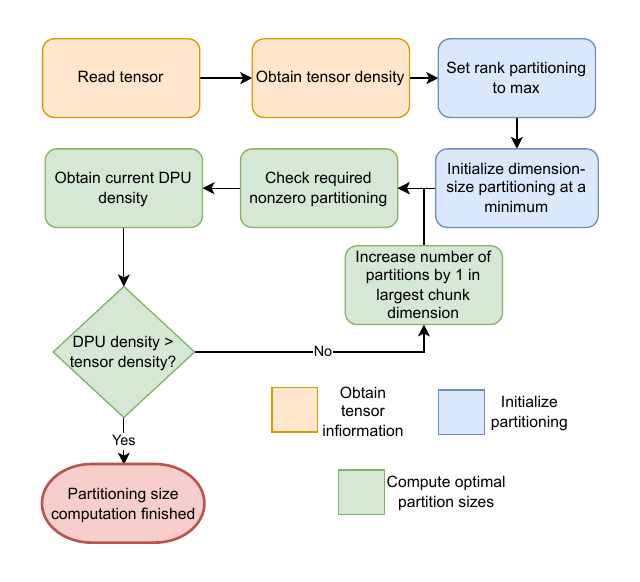}
\caption{Partitioning decider flowchart}
\label{partition_flux}
\vspace{-10pt}
\end{figure}

The proposed approach, represented in Figure \ref{partition_flux}, tackles this issue by using nonzero and dimension size partitions that allow each DPU to mimic the tensor density. Specifically, it measures the tensor density and then iteratively performs dimension size partitioning, calculating the consequent nonzero partitioning required, given the memory size of the DPUs. This process is done by analyzing the impact of each partitioning dimension on DPU memory usage, without requiring explicit partitioning. As the chunk size decreases, less memory is required to store factor matrix elements, which allows the DPUs to store more nonzeros, increasing their density within each DPU. Once the DPU density reaches the tensor density, this process stops. This means that for a perfectly balanced tensor, meaning that the non-zero zeros are evenly distributed across its dimensions, this approach will find the optimal partitioning strategy where it uses the minimum number of chunks and avoids non-zero partitioning. In imbalanced tensors, where the perfect solution does not exist, this algorithm will find a balanced solution, allowing nonzero partitioning to be performed to avoid excessively decreasing the chunk size.\looseness=-1

When the number of DPUs required for spMTTKRP, given by the product between the number of tensor partitions and rank partitions, is larger than available, multiple kernel iterations must be used, each going over different ranks. Since different ranks require different factor matrix elements, the data regarding these must be replaced, eliminating the option of keeping it. However, if the number of tensor partitions alone is greater than the number of available DPUs, each kernel iteration must traverse different tensor chunks, meaning that not even tensor data can be kept between iterations. Since the number of tensor partitions is directly related to dimension size and nonzero partitioning, this provides another reason why these should be minimized.\looseness=-1

\subsection{Fixed-precision and kernel optimizations}

Usually, spMTTKRP is performed using either single-precision or double-precision floating-point arithmetic. However, the DPUs can only perform 32-bit integer addition and 8-bit integer multiplication in hardware, meaning that floating-point operations must be performed by costly software routines, severely hindering performance \cite{UPMEM_SDK, SparseP}. Therefore, replacing floating-precision with fixed-precision will significantly boost this application's performance. To ideally exercise the DPU arithmetic units, the factor matrix elements and tensor nonzero values should be represented using 8-bit integers since these values will be multiplied to obtain a partial result, while the final results of the multiplications can be represented using 32-bit integers since they will only be involved in sum operations. Besides directly affecting the kernel, the data size used will also affect partitioning, since the smaller the data format is, the less memory each element occupies, which decreases the need for partitioning. However, the number of bits used for this data will also affect its precision. The lower this number is, the lower the precision becomes, which can hinder how fast the algorithm will converge (if at all). Extracting the maximum precision for a given number of bits requires knowing the range of values that must be represented to avoid overflow. Therefore, this range has to be determined before precision can be established.\looseness=-1

The range of nonzero values cannot be determined before reading the tensor, which means that the precision attributed to these values must be determined at runtime. The range of matrix values, on the other hand, can be easily determined to be between -1 and 1 by performing a normalization step when computing the factor matrices, assuming an initialization that respects that range. An element ranging between -1 and 1 leads any product between it and another value to have the same range as the latter. This means that intermediate products involving only factor matrix elements will have the same range. Each partial result, given by the multiplication between multiple factor matrix elements and a nonzero value, will have the same range as the latter.\looseness=-1 

Due to the limited precision offered by fixed-point arithmetic, choosing an appropriate normalization norm is very important. As stated in Section \ref{cpd_explanation}, the most commonly used norms are L-2 and L-infinity. However, the L-infinity normalization will, by definition, use the full range of values between -1 and 1, which does not happen for L-2 normalization. Values with higher absolute values suffer less from limited precision, making L-infinity normalization the most indicated for this application.\looseness=-1

When performing multiplications with fixed-precision, the result will have as many integer bits as the sum of the integer bits of both operands, and the same will happen for the decimal bits. However, this result is automatically truncated in the integer part if its size is larger than the output format. In the factor matrix multiplication, since the result ranges from -1 to 1, at least two integer bits need to be kept after truncation. This means that, if both the factor matrix elements and results are represented using 8 bits, the maximum precision of the output is Q2.6. To obtain 6 decimal bits on the output, the input can only contain 3 decimal bits, so the maximum input precision will be Q5.3. For 16-bit data, since the output must be Q2.14, the input becomes Q9.7. For 32-bit data, following the same rationale, the highest input precision achievable is Q17.15. In practice, Q5.3 is too low to reach convergence in the decomposition, which means at least 16 bits must be used to represent factor matrix data and its multiplication results.\looseness=-1

\begin{algorithm}
\small
\caption{Kernel pseudocode}
\label{kernel_pseudocode}
\begin{algorithmic}[1]
\If{$tasklet\_id == 0$}
    \State $load\_contol\_data()$
\EndIf
\State $clean\_output\_buffer\_data()$
\For{nonzero in tasklet\_partition}
    \State $i_1,\dots,i_{n-1},o \gets load\_coordinates(nonzero)$
    \State $x \gets load\_value(nonzero)$
    
\For{$r \in rank\_partition$}
    \State $partial\_result \gets I_1[r][i_1]$
\For{$k=2,\dots,n-1$}
    \State $partial\_result \gets partial\_result * I_k[r][i_k]$
    \State $partial\_result >> matrix\_precision$
\EndFor
\State $partial\_result \gets partial\_result * x$
\State $partial\_result >> (value\_precision+prec\_shift)$
\State $O[r][o] \gets O[r][o] + partial\_result$
\EndFor
\EndFor
\end{algorithmic}
\end{algorithm}

Algorithm \ref{kernel_pseudocode} presents the pseudocode of the UPMEM PIM kernel. The kernel iterates over each nonzero (Lines 5-18) and rank (Lines 8-17), first computing the factor matrix product (Lines 9-12) and then multiplying it by the nonzero value (Line 14) to obtain a partial result. This result is then added to the corresponding output position (Line 16). After each factor matrix element multiplication (Line 11), the precision of the result is re-established by performing as many arithmetic shifts as the number of decimal bits of their format "$matrix\_precision$" (Line 12). The partial results in each DPU are represented using 32-bit integers, and to standardize their format for different tensors, this value is shifted by the number of bits attributed to the tensor data format "$value\_precision$" (Line 15). Additional shifts can be performed to this result, named "$prec\_shift$", to extend the range of representable values within the sum reduction, which avoids overflowing the representation when summing multiple partial results.\looseness=-1

Given that Q17.15 precision is likely to overflow when summing the partial results, we use a $prec\_shift$ value of 3 with this format. The Q9.7 is adopted for 16-bit precision, as it provides the highest precision in 16 bits, and $prec\_shift$ is kept at 0 since the format adopted when transitioning Q9.7 to 32 bits is Q25.7, which has no risk of overflowing. The tensor values are represented using 16-bit integers.\looseness=-1

The pseudocode in Algorithm \ref{kernel_pseudocode} also shows the use of tasklets, as the nonzeros attributed to a DPU are divided among them to allow for a higher throughput of partial results. The number of tasklets used is 16 to extract the maximum throughput of a DPU \cite{UPMEM_SDK}. Using a power of two value also facilitates the workload distribution, since the partition size can be computed with a simple arithmetic shift instead of an integer division. To avoid unnecessary access to the MRAM, which is much slower than access to the WRAM, all the control data needed for the tasklets is passed onto the WRAM before starting the spMTTKRP computation by tasklet 0 (Line 2).\looseness=-1

Additionally, each tasklet has two sequential readers, one for passing the coordinates and another for passing the values of each nonzero to the WRAM before computing its partial result. This optimization extracts the maximum performance out of tensor data accesses because the nonzeros accessed by each tasklet are contiguous. The factor matrix values do not get sent to WRAM because their accesses are unpredictable, and WRAM does not have enough memory to hold all their elements. The partial results of all tasklets are written to a single array, minimizing the memory space they occupy, which increases the memory available for tensor and factor matrix data, reducing dimension size and nonzero partitioning. This approach will, therefore, generate write conflicts. However, CP-ALS is an iterative algorithm that can handle slight imprecision in the spMTTKRP operation. This aspect is already being explored with the usage of fixed-precision, but it is also useful in handling write conflicts. Given that the tensor is sparse, the number of times two tasklets of a DPU will be writing in the same memory position simultaneously will be limited, especially given the faster speed of memory writing provided by PIM. Therefore, it is possible to optimize kernel speed by removing the locking mechanism to handle write conflicts, which will be referred to as locks, as the imprecision caused by it is minor and can be compensated for by the algorithm. Section \ref{Number Format Results}  explores the effect of this optimization with experimental results.\looseness=-1

\subsection{Heterogeneous partitioning}

This section proposes a new design that partitions the spMTTKRP operation between UPMEM PIM and CPU.

When deriving a heterogeneous algorithm, the main goal is to improve the performance of larger workloads, since smaller ones rarely benefit from this approach, as the communication latency easily implies slowdown. Therefore, the mapping strategy adopted in the PIM-only approach is also applied in the heterogeneous one, as it was already optimized for large workloads.\looseness=-1

The workload fraction attributed to CPU and PIM is defined statically, and the workload distributor aims to respect it while attaining the maximum performance from the PIM device. As explained in Section \ref{Partitioning}, rank partitioning does not generate data replication, which means that partitioning the workload through the ranks will not be beneficial, since PIM can handle larger ranks efficiently. When it comes to dimension size and nonzero partitioning, the most efficient partitioning for PIM limits both by having the tensor divided into the least possible number of chunks, each with the highest number of nonzeros that can still fit within a single DPU, avoiding nonzero partitioning. Therefore, to mimic this distribution, the chunks sent to PIM will be from the most to least dense, whose number of elements can still fit within a single DPU.
Then, if the workload fraction sent to PIM is still not enough to respect the desired distribution, chunks that cannot fit in a single DPU start being sent, from least to most dense.\looseness=-1

\section{Experimental results}

\begin{table}[h]
\caption{Characteristics of the used dataset}
\label{tab:tensors}
\centering
\begin{tabular}{|c|c|c|}
\hline
Tensor    & Dimensions                         & Density                           \\ \hline
Nell-2    & 12.1K x 9.2K x 28.8K               & 2.4 x $10^{-5}$  \\ \hline
Nell-1    & 2.9M x 2.1M x 25.5M                & 9.1 x $10^{-13}$ \\ \hline
Amazon & 4.8M x 1.8M x 1.8M & 1.1 x $10^{-10}$ \\ \hline
Delicious & 532.9K x 17.3M x 2.5M x 1.4K       & 4.3 x $10^{-15}$ \\ \hline
LBNL      & 1.6K x 4.2K x 1.6K x 4.2K x 868.1K & 4.2 x $10^{-14}$ \\ \hline
5D\_large      & 1.0M x 100.0K x 300.0K x 400.0K x 50.0K & 1.1 x $10^{-18}$ \\ \hline

\end{tabular}
\end{table}

In this section, PRISM is experimentally evaluated  (including number formats and kernel optimizations) and compared with state-of-the-art implementations on the CPU and GPU. Finally, the heterogeneous implementation is compared to the UPMEM PIM-only and CPU-only. The UPMEM PIM tests are run on the UPMEM PIM server, equipped with a 10-core INTEL Xeon Silver 4210 CPU (base frequency of 2.20 GHz), together with a PIM-enabled DRAM where up to 34 DPU sets were used, corresponding to 2176 DPUs. The CPU tests are also executed on the UPMEM PIM server, using the same CPU. The GPU tests are run on an Nvidia GeForce A100 80GB. The CPU and GPU implementations used both for comparison and on the heterogeneous approach are, respectively, ALTO \cite{ALTO} and BLCO \cite{BLCO}, as they are the state-of-the-art implementations with publicly available code that deliver the highest performance.\looseness=-1

The tensors used for the experimental campaign are presented in Table \ref{tab:tensors}, covering a wide variety of tensor sizes and number of dimensions. Most tensors are obtained from the FROSTT dataset \cite{FROSTT}. The exception is 5D\_large, a tensor specifically created to evaluate the scaling of PRISM for large mode-5 tensors, since the existing tensors in the dataset are too small to test such workloads. This tensor has nonzeros spread randomly over its dimensions, resulting in an overall well-balanced distribution. In this analysis, a tensor is considered small if it cannot occupy most of the processing units when using a common decomposition rank (below 50).\looseness=-1

\subsection{Number Formats and Lock Usage} \label{Number Format Results}

To evaluate how using fixed-precision and removing locks can affect the convergence of the decomposition, five iterations of CP-ALS decomposition were run on the Nell-2, Delicious, and Lbnl tensors, by using single-precision floating-point (Float) and previously elaborated fixed-point formats: Q9.7 (Int7), and Q17.15 with a $prec\_shift$ of 3 (Int15-12). The decomposition for each tensor and each precision was run with a decomposition rank of 10 and the same initialization seed, with and without using locks.\looseness=-1

\begin{figure}[t] 
\centering
\includegraphics[width=\linewidth,trim={0 0 0 0},clip]{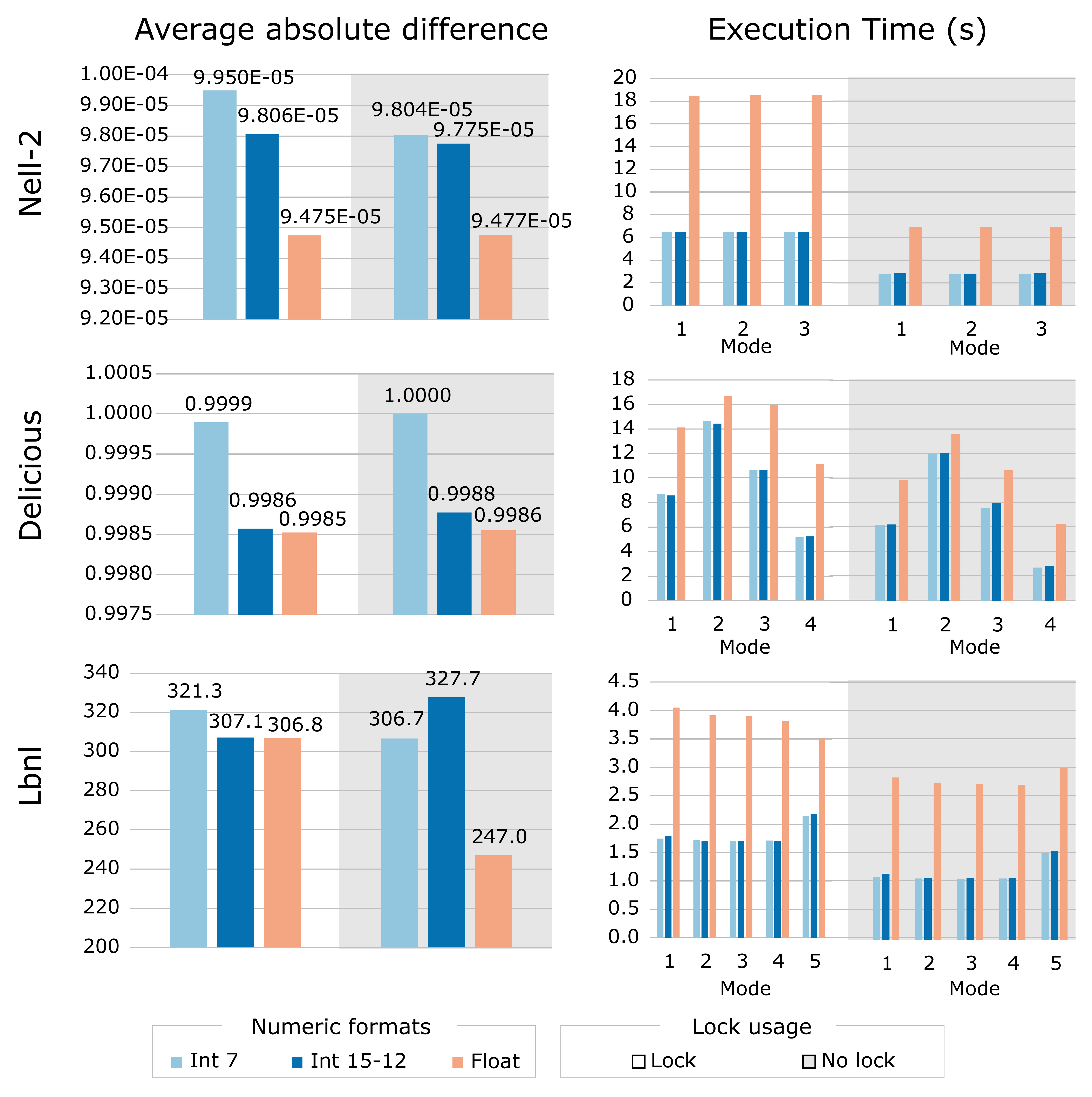}
\caption{Influence of fixed-precision and lock usage on convergence and execution time}
\label{Precisions}
\vspace{-10pt}
\end{figure}

Figure \ref{Precisions} shows the average absolute difference between the decomposition result and the original tensor, allowing for a quantitative analysis of the algorithm's convergence. It also presents the execution time (for each mode) obtained for each tensor in each precision, with and without locks. On Nell-2, the average absolute difference is done for all tensor elements, while for Delicious and Lbnl it is only done for the nonzeros, as the total number of tensor elements is too large to perform this measurement.\looseness=-1
Focusing on the average absolute difference results with lock usage for all three tensors, Int7 presents the highest difference in all cases, while Float presents the lowest. This is to be expected since the single-precision floating-point format has a higher range of representation than any fixed-precision format. However, the difference between the most and least accurate formats is always minimal, with the former achieving a 4.8\%, 0.14\%, and 4.5\% smaller average difference than the latter on Nell-2, Delicious, and Lbnl, respectively. Another important result is that on Nell-2, the difference obtained with Int15-12 is closer to the one obtained with Int7 than with Float, while on the other tensors, the opposite is observed. This is due to the number of factor matrix multiplications increasing with the number of tensor dimensions, generating partial results with lower absolute value, which require increased precision to be represented. This suggests Int15-12 as the preferred format for mode-4 and mode-5 tensors, while both Int7 and Int15-12 can be used for mode-3 tensors.\looseness=-1

Comparing now the results for average absolute difference when removing locks, we see that in some cases, such as Delicious with the Int7 format or Lbnl with the Int15-12 format, it increases, but in other cases, such as Nell-2 with the Int7 format or Lbnl with the Float format, it decreases. This means that the slight imprecision introduced by removing these locking mechanisms does not significantly decrease the algorithm's convergence, having some cases where it can even increase it.\looseness=-1

Regarding the execution times with locks, the results show that both fixed-precision results attain similar performance on all tensors, with the least performant format always being less than 5\% slower than the most performant. Float, on the other hand, is always the slowest format and oscillates between being 13.1\% and 185.5\% slower than the fastest. This means that, as expected, using fixed-precision is significantly faster than floating-point. Similar results obtained with 16-bit integers (Int7) and 32-bit integers (Int15-12) show that the software routines used to perform 16-bit and 32-bit multiplications within the DPU provide similar performance, which means that the main advantage of using 16-bit integers is requiring less partitioning. However, the modest size of the tested tensors makes the kernel computation a significant part of the total execution time, meaning that partitioning the tensor more, which reduces kernel time, can be more beneficial than minimizing data replication. Regarding the usage of locks, kernel times are decreased in all cases from 15.3\% to 62.9\%.\looseness=-1


To analyse how the conclusions regarding the factor matrix data formats translate to larger tensors, 5 iterations of the CP-ALS decomposition were executed on the Nell-1 and Amazon using the Int7 and Float formats with a decomposition rank of 10, the same initialization seed, and with locking mechanisms. 5D\_large was not used in this study because it is an artificial tensor, so its convergence is not relevant to real-world applications.\looseness=-1




The average absolute difference for Nell-1 using the Int7 format was 1.4930, while for Float it was 1.4917. This shows an increase of 0.09\% in average absolute difference from Float to Int7. For Amazon, the average absolute difference with Int7 was 1.3012 and with Float was 1.3001, displaying an even lower increase of 0.08\%. This proves that, even on larger tensors, the usage of fixed-precision does not significantly mitigate the algorithm convergence. This is to be expected, since when using larger tensors, due to the use of L-infinity normalization, the factor matrix elements will have similar absolute values, which means that the precision required for the factor matrix multiplications will be the same. The precision needed to avoid overflowing the DPU partial results is also not affected by the tensor dimension since, due to the partitioning approach used, the DPUs were already being used at full capacity. Using larger tensors will only require the use of more DPUs, which does not affect this overflow.\looseness=-1

The results observed in this section establish Int7 as the preferred format for performing spMTTKRP on mode-3 tensors, while for mode-4 and mode-5 tensors, the Int15-12 format should be used. Regarding lock usage, the decrease in execution time provided by their elimination, coupled with the resulting minimal difference in algorithm convergence, proves that their removal is an effective strategy. Given this, the following tests will be performed with the Int7 format on mode-3 tensors and the Int15-12 format on mode-4 and mode-5 tensors, always without using locks.\looseness=-1

\subsection{Comparison between PIM, CPU, and GPU}

\begin{figure}[t]
\centering
\includegraphics[width=\linewidth,trim={0 0 0 0},clip]{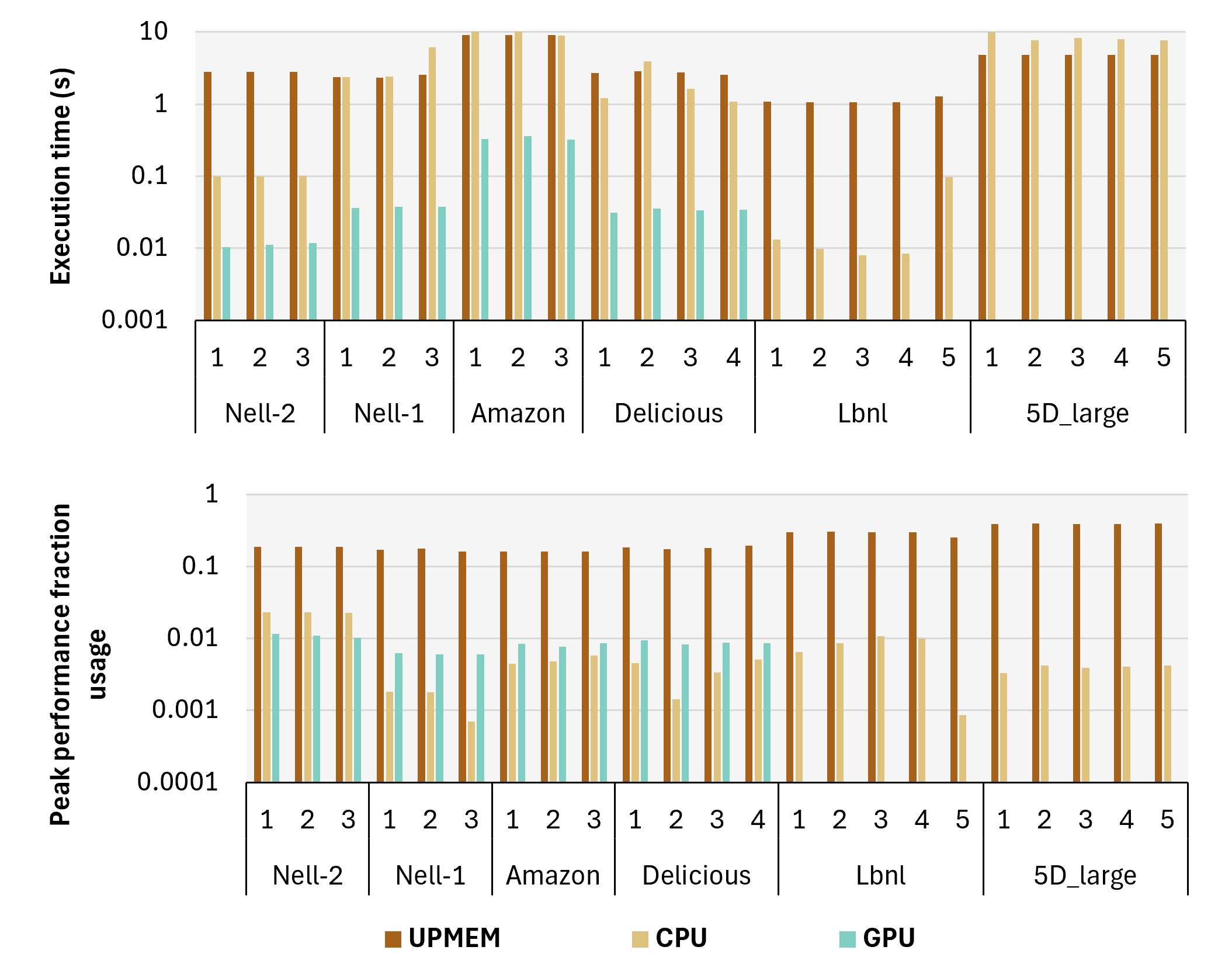}
\caption{spMTTKRP execution time and peak performance fraction usage on UPMEM PIM, CPU, and GPU}
\label{Comparison_fig}
\vspace{-10pt}
\end{figure}

Figure \ref{Comparison_fig} presents the spMTTKRP kernel execution times and peak performance fraction usage on all considered tensors and all modes on UPMEM PIM, CPU, and GPU, using a decomposition rank of 10. Peak performance fraction usage corresponds to the fraction of the peak performance achieved by the device in FLOPS. Since the most recent UPMEM PIM device lacks an accurate method for measuring energy consumption, this becomes the most suitable metric for analyzing efficiency. GPU results are not presented for mode-5 tensors since the BLCO \cite{BLCO} implementation does not support it. The peak performance for CPU and GPU was obtained using \textit{peakperf} \cite{peakperf}, while for the UPMEM PIM device, it is estimated by the product between the peak arithmetic throughput of a DPU obtained using the PrIM benchmark suite \cite{prim} and the number of DPUs used. For tensors that require multiple kernel iterations, the execution time presented is the sum of all iteration times, and the estimated peak performance uses the average number of DPUs per iteration.\looseness=-1

Regarding the execution times, these results show that for the smaller tensors, namely Nell-2 and Lbnl, UPMEM PIM shows worse performance than the CPU. However, on larger tensors, namely Nell-1, Amazon, and 5D\_large, the performance of the UPMEM PIM device surpasses that of the CPU. This shows that, for tensors large enough to utilize most of the UPMEM PIM device's resources, it will outperform the CPU. However, when compared with the kernel execution times on the GPU, the UPMEM PIM results present higher execution times on all tensors. This is a result of the higher processing power of the GPU when compared to the UPMEM PIM device, which is due to the greater amount of computing resources present in the GPU. The price of the devices highlights this discrepancy, with the UPMEM PIM DIMMs being significantly cheaper than most recent accelerators, including the Nvidia A100 GPU \cite{UPMEM_2019,Shen_2024}. This is also reflected in the peak performance fraction usage results, where, even though the GPU presents lower execution times, the peak performance fraction usage is significantly lower, showcasing that its performance is a consequence of raw computing power rather than efficiency.\looseness=-1

The results regarding the peak performance fraction usage on each device show that in all cases, the UPMEM PIM device presents the highest peak performance fraction usage, ranging from 8 to 290 times higher than the CPU and 16 to 30 times higher than the GPU. This demonstrates a much more efficient resource usage in this implementation, which is to be expected due to the fewer costly memory transactions required when using this device.\looseness=-1

\subsection{Heterogeneous CPU + PIM approach}

\begin{figure}[t]
\centering
\includegraphics[width=0.85\linewidth,trim={0 0 0 0},clip]{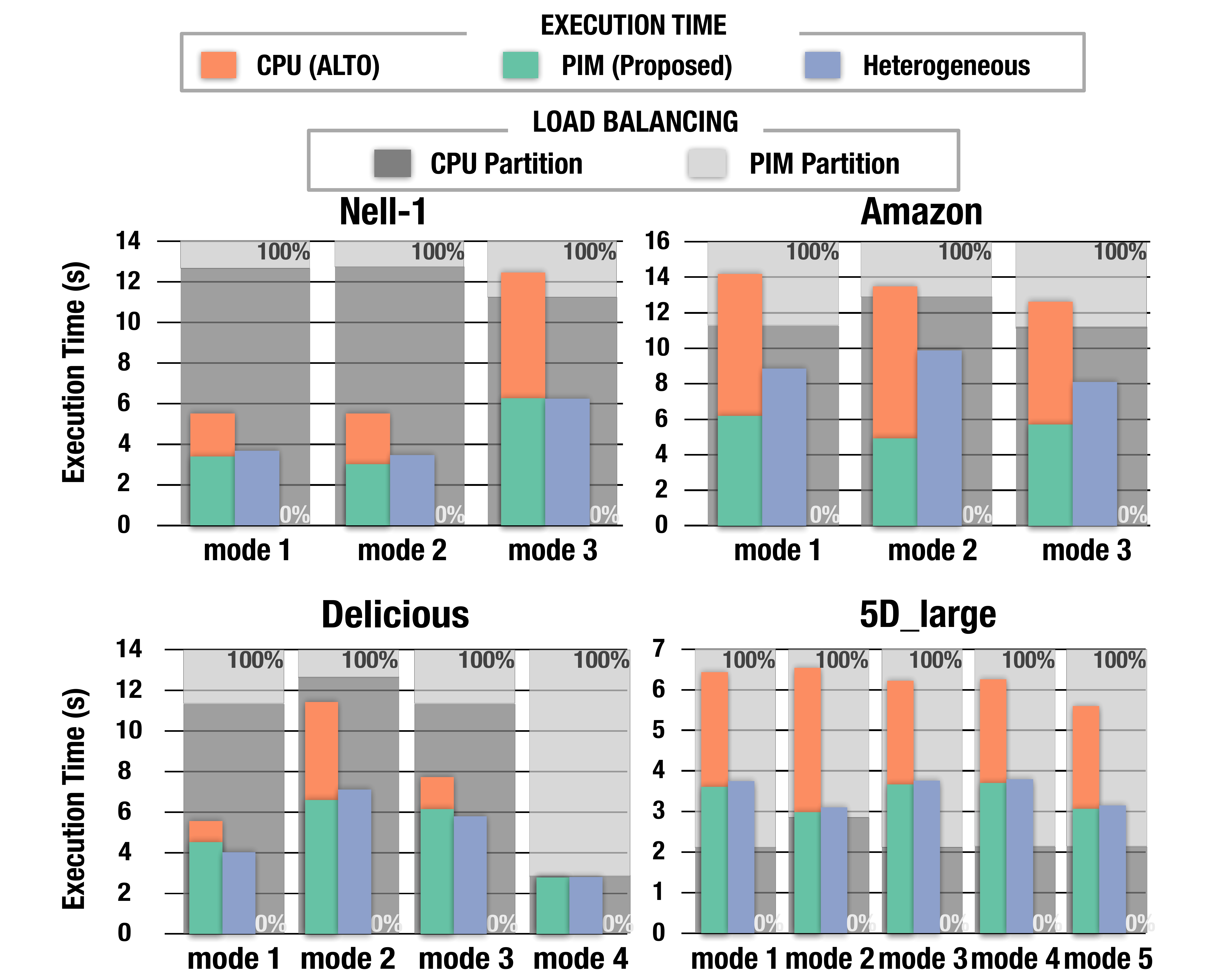}
\caption{Heterogeneous spMTTKRP execution time and workload distribution}
\label{Heterogeneous_times}
\vspace{-10pt}
\end{figure}

\begin{figure}[t]
\centering
\includegraphics[width=\linewidth,trim={0 0cm 0 0},clip]{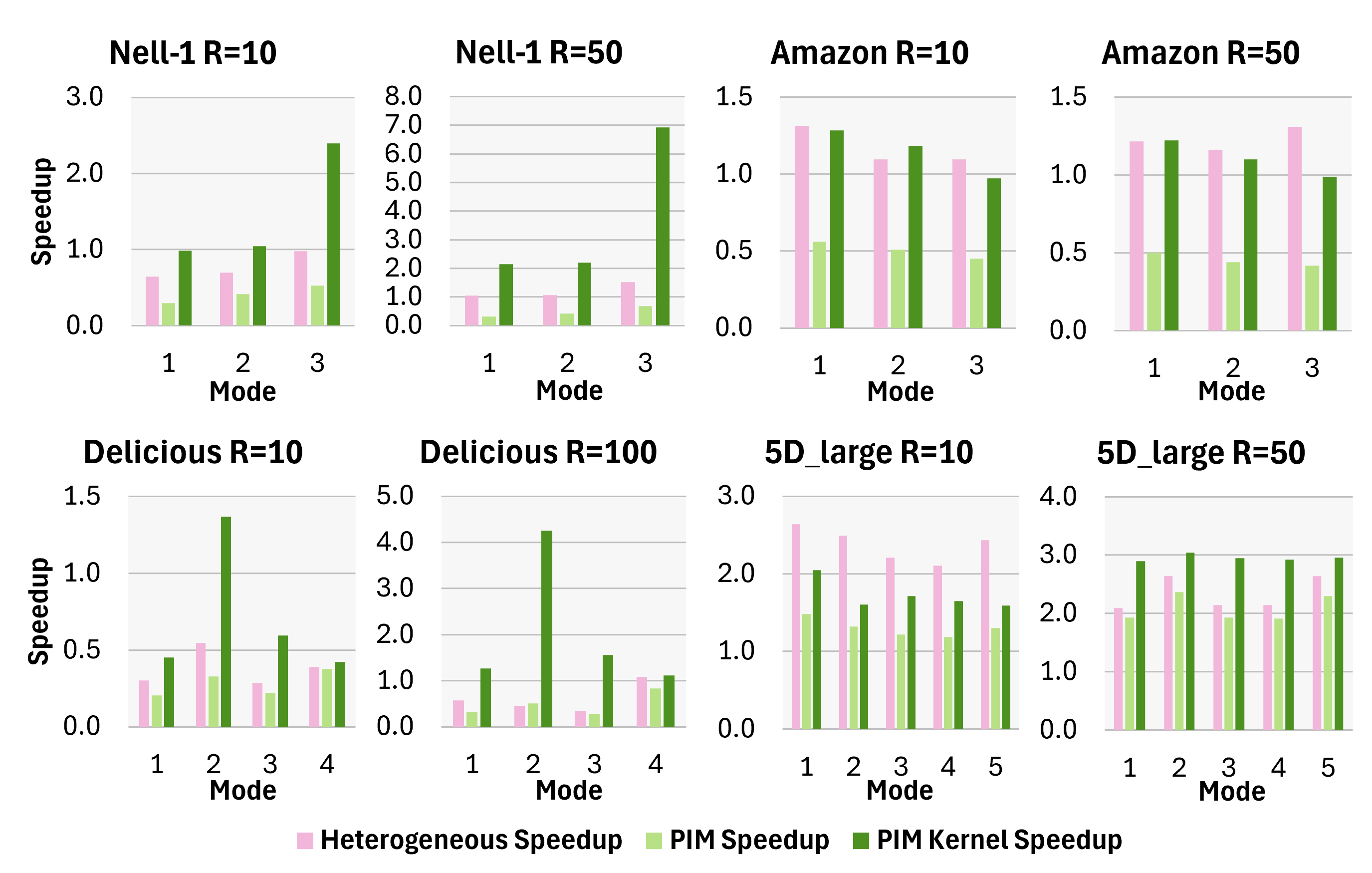}
\caption{Speedup of PIM-only and heterogeneous approaches over ALTO}
\label{Heterogeneous_speedups}
\vspace{-10pt}
\end{figure}

Figure \ref{Heterogeneous_times} presents the evaluation of the proposed heterogeneous spMTTKRP CPU+PIM approach for the largest tensors (Nell-1, Amazon, Delicious, and 5D\_large), with a decomposition rank of 10. The results show the full execution time for the heterogeneous approach (blue bar), as well as the workload distribution (gray bars) and the time taken for PIM (green) and ALTO (red) to perform their partitions separately. Figure \ref{Heterogeneous_speedups} shows the speedups of the PIM-only (green bars) and the Heterogeneous (pink bar) implementations over ALTO. For the PIM-only implementation, two speedup values are shown, one that only considers the kernel time (dark green bar), and another that also considers memory transactions and sum reduction (light green bar). The decomposition ranks used are R=10 and higher, aiming to create a workload large enough to use all processing units. The comparison is, once again, done with ALTO since it is the most performant state-of-the-art CPU implementation publicly available, allowing it to be run in the same system and tensors as the PIM-only and heterogeneous implementations. The CPU partition of the heterogeneous approach is also performed using ALTO.\looseness=-1

Figure \ref{Heterogeneous_times} shows that the execution times of the CPU and UPMEM PIM on Nell-1, Amazon, and 5D\_large for PIM and ALTO are very similar, even though on Amazon and Nell-1, at least 70\% of the workload is assigned to ALTO. However, in 5D\_large, ALTO is only performing between 30\% and 40\% of the workload. This improvement is also visible in the speedups of Figure \ref{Heterogeneous_speedups}, where, even though Nell-1 on rank 10 still does not reach speedup, Amazon and 5D\_large reach it in all modes with the heterogeneous approach. Note that 5D\_large reaches it using only PIM, even considering memory transactions and sum reduction, showcasing the much better performance of PIM on larger workloads. Figure \ref{Heterogeneous_times} also shows that the sum of the execution times of ALTO and PIM on Delicious mode 4 is similar to the heterogeneous time. However, in all other cases, the time for the heterogeneous approach is 1.4 to 2.1 times smaller than the sum, confirming that this approach extracts the most performance from these devices.\looseness=-1

When comparing the results with different decomposition ranks in Figure \ref{Heterogeneous_speedups}, almost all cases present better speedups for higher ranks. However, on Amazon, which is the largest tensor, the speedups barely suffer any variation when increasing the rank. This means that the workload provided by this tensor is sufficient to use all the DPUs for a rank 10 decomposition, maximizing the obtainable speedup. The speedup results for the Delicious tensor on mode 2 of the heterogeneous approach also show an exception to the usual increase in performance on higher decomposition ranks. Yet, this can be explained by the large sum-reduction required by this spMTTKRP iteration, since the second dimension of Delicious is the largest of any tensor, coupled with the decreased performance of the proposed implementation on this tensor due to its imbalanced nature. This substantially increases the workload of the CPU and decreases the benefit gained from the PIM contribution, mitigating the benefits of the heterogeneous approach. On the other hand, it also leads to a higher speedup when only kernel time is considered, proving the significant performance impact of the sum reduction.\looseness=-1

The results of the PIM Kernel Speedup on Figure \ref{Heterogeneous_speedups} show an average speedup of 1.29 and 2.50 for low ranks and high ranks, respectively. On the other hand, the complete operation in PIM only shows an average speedup of 0.69 and 1.01 for low ranks and high ranks, respectively. This showcases the performance impact of the memory transactions and sum reduction performed, highlighting the importance of the approach devised in Section \ref{Partitioning} to minimize data replication and, consequently, the time taken by these operations. Even with these optimizations, on average, 52\% of the execution time of the PIM-only approach is still attributed to them. This presents the drawbacks of implementing the MTTKRP operation on a distributed memory system, where data replication is required to provide the necessary information to all processing units, and their results must be reduced. Nonetheless, even with an approach that aims to reduce data replication instead of kernel time, significant kernel speedup is obtained, especially on higher ranks, demonstrating the benefits of the PIM architecture on memory-bound workloads.\looseness=-1

Overall, the experimental results show that the heterogeneous approach designed in this work is capable of outperforming the state-of-the-art CPU implementations, especially for large workloads. This is achieved by performing multiple optimizations and reducing the data replication with the proposed algorithm.\looseness=-1

\section{Conclusion}

This work proposed and characterized PRISM, a novel approach to the implementation of spMTTKRP using UPMEM PIM for CP-ALS. It considered multiple data formats, partitioning techniques, kernel optimizations, and a heterogeneous CPU + UPMEM PIM approach. The results showcased high efficiency and significant performance on large workloads and well-balanced tensors, highlighting the capabilities of PIM technology. They also displayed limitations that stem from the distributed memory system.
To this respect, this work also launches a new challenge for the future of this promising technology by posing the question of how much spMTTKRP and other operations with similar sorts of dependencies could benefit from the usage of a PIM device with global memory accessible for all DPUs, or with the capability of direct inter-DPU communication.\looseness=-1

\section*{Acknowledgments}
This work was supported by FCT (Fundação para a Ciência e a Tecnologia, Portugal) and EuroHPC Joint Undertaking through the UIDB/50021/2020 project and grant agreements No 101092877 (SYCLOPS) and No 101202459 (DARE SGA1 project).\looseness=-1

\balance
\bibliographystyle{IEEEtranS}
\bibliography{reference}

@inproceedings{nlp,
author = {Kang, U. and Papalexakis, Evangelos and Harpale, Abhay and Faloutsos, Christos},
title = {GigaTensor: scaling tensor analysis up by 100 times - algorithms and discoveries},
year = {2012},
isbn = {9781450314626},
booktitle = {Proceedings of the 18th ACM SIGKDD International Conference on Knowledge Discovery and Data Mining}
}

@article{CP,
  title={Foundations of the PARAFAC procedure: Models and conditions for an “explanatory” multi-modal factor analysis},
  author={Harshman, Richard A and others},
  journal={UCLA working papers in phonetics},
  volume={16},
  number={1},
  pages={84},
  year={1970},
  publisher={Los Angeles, CA}
}

@article{parti_paper,
  title={Scalable tensor decompositions in high performance computing environments},
  author={Li, Jiajia},
  year={2018},
  publisher={Georgia Institute of Technology}
}

@INPROCEEDINGS{HiCOO,
  author={Li, Jiajia and Sun, Jimeng and Vuduc, Richard},
  booktitle={SC18: International Conference for High Performance Computing, Networking, Storage and Analysis}, 
  title={HiCOO: Hierarchical Storage of Sparse Tensors}, 
  year={2018},
  volume={},
  number={},
  pages={238-252},
  doi={10.1109/SC.2018.00022}}

@inproceedings{ALTO,
author = {Helal, Ahmed E. and Laukemann, Jan and Checconi, Fabio and Tithi, Jesmin Jahan and Ranadive, Teresa and Petrini, Fabrizio and Choi, Jeewhan},
title = {ALTO: adaptive linearized storage of sparse tensors},
year = {2021},
isbn = {9781450383356},
publisher = {Association for Computing Machinery},
address = {New York, NY, USA},
url = {https://doi.org/10.1145/3447818.3461703},
doi = {10.1145/3447818.3461703},
booktitle = {Proceedings of the 35th ACM International Conference on Supercomputing},
pages = {404–416},
numpages = {13},
location = {Virtual Event, USA},
series = {ICS '21}
}

@inproceedings{FLYCOO_FPGA,
author = {Wijeratne, Sasindu and Wang, Ta-Yang and Kannan, Rajgopal and Prasanna, Viktor},
title = {Accelerating Sparse MTTKRP for Tensor Decomposition on FPGA},
year = {2023},
isbn = {9781450394178},
publisher = {Association for Computing Machinery},
address = {New York, NY, USA},
url = {https://doi.org/10.1145/3543622.3573179},
doi = {10.1145/3543622.3573179},
booktitle = {Proceedings of the 2023 ACM/SIGDA International Symposium on Field Programmable Gate Arrays},
pages = {259–269},
numpages = {11},
location = {Monterey, CA, USA},
series = {FPGA '23}
}

@INPROCEEDINGS{Dynasor,
  author={Wijeratne, Sasindu and Kannan, Rajgopal and Prasanna, Viktor},
  booktitle={2023 IEEE 35th International Symposium on Computer Architecture and High Performance Computing (SBAC-PAD)}, 
  title={Dynasor: A Dynamic Memory Layout for Accelerating Sparse MTTKRP for Tensor Decomposition on Multi-core CPU}, 
  year={2023},
  volume={},
  number={},
  pages={23-33},
  doi={10.1109/SBAC-PAD59825.2023.00012}}

@inproceedings{GPU_spMTTKRP,
author = {Wijeratne, Sasindu and Kannan, Rajgopal and Prasanna, Viktor},
title = {Sparse MTTKRP Acceleration for Tensor Decomposition on GPU},
year = {2024},
isbn = {9798400705977},
publisher = {Association for Computing Machinery},
address = {New York, NY, USA},
url = {https://doi.org/10.1145/3649153.3649187},
doi = {10.1145/3649153.3649187},
booktitle = {Proceedings of the 21st ACM International Conference on Computing Frontiers},
pages = {88–96},
numpages = {9},
location = {Ischia, Italy},
series = {CF '24}
}

@inproceedings{BLCO,
author = {Nguyen, Andy and Helal, Ahmed E. and Checconi, Fabio and Laukemann, Jan and Tithi, Jesmin Jahan and Soh, Yongseok and Ranadive, Teresa and Petrini, Fabrizio and Choi, Jee W.},
title = {Efficient, out-of-memory sparse MTTKRP on massively parallel architectures},
year = {2022},
isbn = {9781450392815},
publisher = {Association for Computing Machinery},
address = {New York, NY, USA},
url = {https://doi.org/10.1145/3524059.3532363},
doi = {10.1145/3524059.3532363},
booktitle = {Proceedings of the 36th ACM International Conference on Supercomputing},
articleno = {26},
numpages = {13},
location = {Virtual Event},
series = {ICS '22}
}

@article{SparseP,
author = {Giannoula, Christina and Fernandez, Ivan and Luna, Juan G\'{o}mez and Koziris, Nectarios and Goumas, Georgios and Mutlu, Onur},
title = {SparseP: Towards Efficient Sparse Matrix Vector Multiplication on Real Processing-In-Memory Architectures},
year = {2022},
issue_date = {March 2022},
publisher = {Association for Computing Machinery},
address = {New York, NY, USA},
volume = {6},
number = {1},
url = {https://doi.org/10.1145/3508041},
doi = {10.1145/3508041},
journal = {Proc. ACM Meas. Anal. Comput. Syst.},
month = feb,
articleno = {21},
numpages = {49}
}

@article{WFA,
  title={High-throughput pairwise alignment with the wavefront algorithm using processing-in-memory},
  author={Diab, Safaa and Nassereldine, Amir and Alser, Mohammed and Luna, Juan G{\'o}mez and Mutlu, Onur and Hajj, Izzat El},
  journal={arXiv preprint arXiv:2204.02085},
  year={2022}
}

@article{sa,
    author = {Diab, Safaa and Nassereldine, Amir and Alser, Mohammed and Gómez Luna, Juan and Mutlu, Onur and El Hajj, Izzat},
    title = {A framework for high-throughput sequence alignment using real processing-in-memory systems},
    journal = {Bioinformatics},
    volume = {39},
    number = {5},
    pages = {btad155},
    year = {2023},
    month = {03},
    issn = {1367-4811},
    doi = {10.1093/bioinformatics/btad155},
    url = {https://doi.org/10.1093/bioinformatics/btad155},
    eprint = {https://academic.oup.com/bioinformatics/article-pdf/39/5/btad155/50204881/btad155.pdf},
}

@article{PIM_Join,
author = {Lim, Chaemin and Lee, Suhyun and Choi, Jinwoo and Lee, Jounghoo and Park, Seongyeon and Kim, Hanjun and Lee, Jinho and Kim, Youngsok},
title = {Design and Analysis of a Processing-in-DIMM Join Algorithm: A Case Study with UPMEM DIMMs},
year = {2023},
issue_date = {June 2023},
publisher = {Association for Computing Machinery},
address = {New York, NY, USA},
volume = {1},
number = {2},
url = {https://doi.org/10.1145/3589258},
doi = {10.1145/3589258},
journal = {Proc. ACM Manag. Data},
month = jun,
articleno = {113},
numpages = {27}
}

@INPROCEEDINGS{rna_pim,
  author={Chen, Liang-Chi and Ho, Chien-Chung and Chang, Yuan-Hao},
  booktitle={2023 60th ACM/IEEE Design Automation Conference (DAC)}, 
  title={UpPipe: A Novel Pipeline Management on In-Memory Processors for RNA-seq Quantification}, 
  year={2023},
  volume={},
  number={},
  pages={1-6},
  doi={10.1109/DAC56929.2023.10247915}}

@inproceedings{CP_ALS2,
 author = {Choi, Joon Hee and Vishwanathan, S.},
 booktitle = {Advances in Neural Information Processing Systems},
 editor = {Z. Ghahramani and M. Welling and C. Cortes and N. Lawrence and K.Q. Weinberger},
 pages = {},
 publisher = {Curran Associates, Inc.},
 title = {DFacTo: Distributed Factorization of Tensors},
 volume = {27},
 year = {2014}
}

@misc{UPMEM_SDK, title={Coding tips and recommended practices}, url={https://sdk.upmem.com/2025.1.0/fff\_CodingTips.html}, journal={Coding tips and recommended practices - UPMEM DPU SDK 2025.1.0 Documentation}, author={UPMEM}, year={2025}}

@online{FROSTT,
  title = {{FROSTT}: The Formidable Repository of Open Sparse Tensors and Tools},
  author = {Smith, Shaden and Choi, Jee W. and Li, Jiajia and Vuduc, Richard and Park, Jongsoo and Liu, Xing and Karypis, George},
  url = {http://frostt.io/},
  year = {2017},
}

@misc{peakperf,
	author = {Dr-Noob},
	title = {{G}it{H}ub - {D}r-{N}oob/peakperf: {A}chieve peak performance on x86 {C}{P}{U}s and {N}{V}{I}{D}{I}{A} {G}{P}{U}s},
	howpublished = {\url{https://github.com/Dr-Noob/peakperf}},
	year = {2021},
	note = {[Accessed 02-02-2025]},
}

@inproceedings{prim,
  title={{Benchmarking Memory-centric Computing Systems: Analysis of Real Processing-in-Memory Hardware}},
  author={Juan Gómez-Luna and Izzat El Hajj and Ivan Fernandez and Christina Giannoula and Geraldo F. Oliveira and Onur Mutlu},
  booktitle={2021 12th International Green and Sustainable Computing Conference (IGSC)},
  year={2021},
  organization={IEEE}
}

@misc{UPMEM_2019, title={The true processing in memory accelerator}, url={https://old.hotchips.org/hc31/HC31\_1.4\_UPMEM.FabriceDevaux.v2\_1.pdf}, journal={Best performance and efficiency for big data & AI}, author={UPMEM}, year={2019}}

@misc{Shen_2024, title={How much is an nvidia A100?}, url={https://modal.com/blog/nvidia-a100-price-article}, journal={Modal}, author={Shen, Margaret}, year={2024}}

@article{sp,
author = {Anandkumar, Animashree and Ge, Rong and Hsu, Daniel and Kakade, Sham M. and Telgarsky, Matus},
title = {Tensor decompositions for learning latent variable models},
year = {2014},
issue_date = {January 2014},
publisher = {JMLR.org},
volume = {15},
number = {1},
issn = {1532-4435},
journal = {J. Mach. Learn. Res.},
month = jan,
pages = {2773–2832},
numpages = {60}
}

@INPROCEEDINGS{cp2,
  author={Chen, Yuedan and Xiao, Guoqing and Özsu, M. Tamer and Tang, Zhuo and Zomaya, Albert Y. and Li, Kenli},
  booktitle={2022 IEEE 38th International Conference on Data Engineering (ICDE)}, 
  title={Exploiting Hierarchical Parallelism and Reusability in Tensor Kernel Processing on Heterogeneous HPC Systems}, 
  year={2022},
  volume={},
  number={},
  pages={2522-2535},
  doi={10.1109/ICDE53745.2022.00234}}

@article{CP_ALS3,
title = {Parallel algorithms for tensor completion in the CP format},
journal = {Parallel Computing},
volume = {57},
pages = {222-234},
year = {2016},
issn = {0167-8191},
doi = {https://doi.org/10.1016/j.parco.2015.10.002},
url = {https://www.sciencedirect.com/science/article/pii/S0167819115001210},
author = {Lars Karlsson and Daniel Kressner and André Uschmajew}
}

@INPROCEEDINGS{cp1,
  author={Kurt, Süreyya Emre and Raje, Saurabh and Sukumaran-Rajam, Aravind and Sadayappan, P.},
  booktitle={2022 IEEE International Parallel and Distributed Processing Symposium (IPDPS)}, 
  title={Sparsity-Aware Tensor Decomposition}, 
  year={2022},
  volume={},
  number={},
  pages={952-962},
  doi={10.1109/IPDPS53621.2022.00097}}

@article{Z-Morton,
  title={A computer oriented geodetic data base and a new technique in file sequencing},
  author={Morton, Guy M},
  year={1966},
  publisher={International Business Machines Company New York}
}

@ARTICLE{CP_ALS,
  author={Vorobyov, S.A. and Yue Rong and Sidiropoulos, N.D. and Gershman, A.B.},
  journal={IEEE Transactions on Signal Processing}, 
  title={Robust iterative fitting of multilinear models}, 
  year={2005},
  volume={53},
  number={8},
  pages={2678-2689},
  doi={10.1109/TSP.2005.850343}}

\end{document}